\documentclass{emulateapj}

\newcommand{\xtej}{XTE J1814$-$338}

\begin{document}

\title{The X-Ray Position and Optical Counterpart of the Accretion-Powered Millisecond Pulsar \xtej}

\author{Miriam I. Krauss, Zhongxiang Wang, Allyn Dullighan\altaffilmark{1}, Adrienne M. Juett\altaffilmark{2}, David L. Kaplan, and Deepto Chakrabarty}
\affil{Department of Physics and Center for Space Research, Massachusetts Institute of Technology,
Cambridge, MA 02139 \\
miriam, wangzx, allyn, ajuett, dlk, deepto@space.mit.edu}
\author{Marten H. van Kerkwijk}
\affil{Department of Astronomy and Astrophysics, University of Toronto, 60 St. George Street, Toronto, ON M5S 3H8, Canada\\
mhvk@astro.utoronto.ca}
\author{Danny Steeghs and Peter G. Jonker\altaffilmark{3,4}}
\affil{Harvard-Smithsonian Center for Astrophysics, 60 Garden St., Cambridge, MA 02138\\
dsteeghs@cfa.harvard.edu, p.jonker@sron.nl}
\author{and \\
\vspace{3mm}
Craig B. Markwardt\altaffilmark{5}}
\affil{Department of Astronomy, University of Maryland,
College Park, MD 20742 \\
craigm@milkyway.gsfc.nasa.gov}

\altaffiltext{1}{Current address: Lincoln Laboratory, Massachusetts Institute of Technology, Lexington, MA 02420}
\altaffiltext{2}{Current address: Department of Astronomy, University of Virginia, Charlottesville, VA 22903}
\altaffiltext{3}{Chandra Fellow}
\altaffiltext{4}{Current address: SRON National Institute for Space Research, Sorbonnelaan 2, 3584 CA Utrecht, The Netherlands}
\altaffiltext{5}{Also Laboratory for High Energy Astrophysics, NASA Goddard Space Flight Center, Greenbelt, MD 20771}


\begin{abstract}

We report the precise optical and X-ray localization of the 3.2~ms accretion-powered X-ray
pulsar \xtej~with data from the {\em Chandra X-Ray Observatory} as well as optical observations conducted during the 2003 June discovery outburst.  Optical
imaging of the field during the outburst of this soft X-ray transient reveals an $R=18$ star at the
X-ray position.  This star is absent ($R>20$) from an archival 1989 image
of the field and brightened during the 2003 outburst, and we therefore identify
it as the optical counterpart of \xtej.  The best source position derived from optical astrometry is R.A. = $\rm18^{h}13^{m}39$\fs04, Dec.= $-$33$^{\circ}$46\arcmin22\farcs3 (J2000).  The featureless X-ray spectrum of the pulsar in outburst is best fit by an absorbed power-law (with photon index $\gamma=1.41\pm 0.06$) plus blackbody (with $kT=0.95\pm 0.13$ keV) model, where the blackbody component contributes approximately 10\% of the source flux.  The optical broad-band spectrum shows evidence for an excess of infrared emission with respect to an X-ray heated accretion disk model, suggesting a significant contribution from the secondary or from a synchrotron-emitting region.  A follow-up observation performed when \xtej~was in quiescence reveals no counterpart to a limiting magnitude of $R = 23.3$.  This suggests that the secondary is an M3 V or later-type star, and therefore very unlikely to be responsible for the soft excess, making synchroton emission a more reasonable candidate.

\end{abstract}

\section{Introduction}

It has long been believed that millisecond radio pulsars are the spun-up products of sustained mass transfer onto neutron stars in low-mass X-ray binaries \citep[e.g.,][]{bhattacharya91}.  Their presumed immediate progenitors, accretion-powered millisecond X-ray pulsars, proved elusive for many years, but six such systems are now known: SAX J1808.4$-$3658 \citep{wijnands98,chakrabarty98}; XTE J1751$-$305 \citep{markwardt02}; XTE J0929$-$314 \citep{galloway02}; XTE J1807$-$294 \citep{markwardt03}; XTE J1814$-$314 \citep{markwardt03b,strohmayer03}; and IGR J00291+5934 \citep{galloway05}.  In addition, 13 accreting neutron stars also show millisecond oscillations during thermonuclear X-ray bursts \citep[see][for a review]{strohmayer04}.  These systems are now understood as nuclear-powered millisecond pulsars, with the burst oscillations tracing the pulsar spin \citep{strohmayer02,chakrabarty03}.  

The soft X-ray transient XTE J1814$-$314 ($l$ = 358\fdg7, $b$ = $-$7\fdg6) was discovered in outburst on 2003 June 5 during scans of the central Galactic plane with the {\em Rossi X-Ray Timing Explorer} \citep[{\it RXTE};][]{markwardt03b}.  The outburst lasted for approximately 55 days, and had a peak 2--10 keV flux of around 13 mCrab.  {\em RXTE} observations also established the source as a 314 Hz (3.2 ms) accretion-powered X-ray pulsar \citep{markwardt03b} in a 4.3~hour binary with a main sequence companion of at least 0.17~$M_\odot$ \citep[using the mass function of $0.002016~M_{\sun}$ in][assuming a neutron star mass of $1.4~M_{\sun}$]{markwardt03c}.  Over two dozen thermonuclear X-ray bursts with millisecond oscillations at the spin frequency were detected from XTE J1814$-$338 during the 2003 June outburst \citep{strohmayer03}.  These burst oscillations had the particularly interesting characteristic of containing significant harmonic content, which allowed \citet{bhattacharyya05} to constrain the neutron star as well as orbital parameters.  One of these bursts showed evidence for photospheric radius expansion, allowing \citet{strohmayer03} to infer a source distance of $8.0\pm 1.6$~kpc.\footnote{\scriptsize{This distance estimate should be viewed with some caution, since the burst it is based on has some features which suggest it may not have been Eddington-limited.  If the burst was not actually Eddington limited, $L_{\rm burst} < L_{\rm Edd}$, and the distance to \xtej~may be less than $8.0 \pm 1.6$ kpc. We therefore consider 8.0 kpc to be an upper limit for the source distance.}}

We obtained a brief observation of the source on 2003 June 20 with the {\em Chandra X-Ray Observatory} for the purpose of measuring its position, and we used this position to identify the optical counterpart \citep{krauss03}.  Subsequent optical spectroscopy revealed strong emission lines of H and He including double-peaked H$\alpha$ emission, indicative of an interacting binary \citep{steeghs03}.   In this paper, we present a detailed report on our {\em Chandra} and optical observations of \xtej.    In Section~\ref{xopt}, we present X-ray and optical imaging and the precise localization and flux measurements of the optical counterpart.  We analyze the X-ray spectrum in Section~\ref{xspec}, and in Section~\ref{disc} we discuss the implications of the current data in our understanding of the physical parameters of \xtej.

\section{X-Ray and Optical Imaging} \label{xopt}

We observed \xtej~with {\it Chandra} for 9.7 ks on 2003 June 20 using the High Energy Transmission Grating Spectrometer (HETGS) with the spectroscopic array of the Advanced CCD Imaging Spectrometer (ACIS-S).  The HETG is comprised of two sets of transmission gratings:  the Medium Energy Gratings (MEGs), with a range of 2.5--31 \AA~(0.4--5.0 keV), and the High Energy Gratings (HEGs), with a range of 1.2--15 \AA~(0.8--10 keV).  The HETGS spectra are imaged by ACIS-S, an array of six CCD detectors.  The HETG/ACIS-S combination provides both an undispersed (zeroth-order) image as well as dispersed spectra from the gratings.  The spatially overlapping spectral orders are sorted using the intrinsic energy resolution of the ACIS-S CCDs.  The first-order MEG (HEG) spectrum has a spectral resolution of $\Delta\lambda=0.023$ (0.012) \AA~FWHM.

All data processing was done with the CIAO analysis package\footnote{http://cxc.harvard.edu/ciao/} (version 2.3).  We summed the dispersed first-order events in 500 s time bins to create an X-ray lightcurve, which we searched for signs of orbital modulation.  We note that the observation spanned only 60\% of the 4.3 hour orbital period, but did cover the time period during which the neutron star was behind the secondary.  The data were consistent with a constant count rate of $\rm4.6\pm0.1~counts~s^{-1}$, and we did not detect any evidence of an X-ray eclipse.

We corrected the observation aspect to be consistent with the calibration available as of 2004 March 22.  No sources other than \xtej~were detected in the field.  Due to the high source count rate, the zeroth order image is over 75\% piled up, which suppresses counts in the image core and results in a characteristic ``doughnut'' shaped point spread function \citep[see, e.g.,][]{davis01}.  However, this did not affect our ability to use the zeroth-order image to obtain a precise positional measurement with the CIAO tool {\tt wavdetect}.  In order to account for the pileup, we used large wavelet scales (8 and 12 pixels), which are not sensitive to the core of the point spread function.  The best-fit X-ray position of \xtej~was R.A. = $\rm18^{h}13^{m}39$\fs02, Dec. = $-$33$^{\circ}$46\arcmin22\farcs3 (equinox J2000.0) with a 90\% confidence radius of 0\farcs6.

\begin{figure}
 \resizebox{!}{2.7in}{\includegraphics{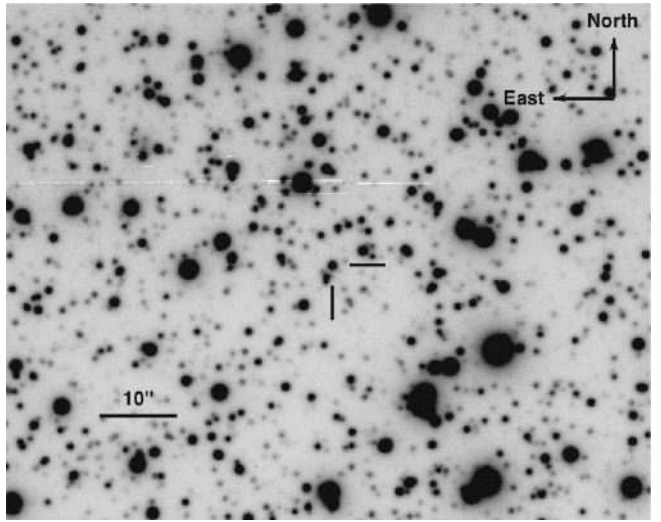}}
           \caption{$I$-band MagIC image of the XTE J1814$-$338 field from 2003 June 21.  The $I = 17.4$ optical counterpart is marked near the center of the image.  North is up and east is to the left.}
           \label{oc}
\end{figure}

The precise X-ray source position facilitated the identification of an optical counterpart.  We obtained $BVR$ images (Harris broadband filter set, which closely approximates Johnson-Cousins) of the \xtej~field on 2003 June 6 and $BR$ images on 2003 June 7 using the LDSS2 camera on the 6.5-m Magellan/Clay telescope at the Las Campanas Observatory in Chile (field of view of $2.5\arcmin \times 2.5\arcmin$ with a scale of $0.38\arcsec$/pixel; the seeing on both nights was approximately $0.8\arcsec$).  All frames were de-biased and flat-fielded using the IRAF package.  An astrometric solution was derived using the USNO-B1.0 catalog \citep{monet03}, giving a standard deviation of 0\farcs35.  Using this solution, we find that there is one optical source within the {\it Chandra} error circle.  This source is not present in a 1989 Digitized Sky Survey\footnote{http://archive.stsci.edu/dss/index.html} image to a limiting magnitude of $R \sim 20$: we therefore identify this source as the optical counterpart.  Further refinement of the optical position is described below.  Flux calibration for the first night was done using the photometric standard star Mark~A \citep{landolt92}, and calibration for the second night was made by fitting several in-field stars to the previous night's measurements; these values are shown in Table~\ref{mag}.  Note that these fluxes replace the incorrectly calibrated values reported in \citet{krauss03}.  The optical counterpart brightened slightly over the course of the two days, paralleling the increase in X-ray flux over the same time period as measured by {\it RXTE}.

\begin{deluxetable}{llcccc}
\tabletypesize{\footnotesize}
\tablewidth{0pt}
\tablecaption{Optical Magnitudes of \xtej}
\tablehead{
\multicolumn{2}{c}{Observation time (UT)} & \colhead{$B$} & \colhead{$V$} & \colhead{$R$} & \colhead{$I$}}
\startdata
2003 June 6\tablenotemark{a} & 10:25 & 18.96 & 18.59 & 18.26 & \nodata \\
2003 June 7\tablenotemark{a}  & 09:57 & 18.71 & \nodata & 18.05 & \nodata \\
2003 June 21\tablenotemark{b} & 03:32 & 18.77 & 18.48 & 18.33 & 17.47 \\
                                                          & 05:06 & 18.64 & 18.39 & 18.16 & 17.37 \\
                                                          & 05:13 & 18.62 & 18.34 & 18.21 & 17.42 \\
                                                          & 06:26 & 18.61 & 18.33 & 18.16 & 17.35 \\
2003 June 24\tablenotemark{b} & 06:39 & 18.75 & 18.48 & 18.29 & 17.38 \\
2004 March 15 & 07:47 & \nodata & \nodata & $> 23.3$\tablenotemark{c} & \nodata \\
\enddata
\label{mag}
\tablenotetext{a}{Calibration errors are $\pm$0.06 magnitudes; relative uncertainty between these two nights is $\pm$0.01 magnitudes.}
\tablenotetext{b}{Calibration errors are $\pm$0.02 magnitudes.}
\tablenotetext{c}{\xtej~was not detected; this value is the $3\sigma$ limiting magnitude.}
\end{deluxetable}

On 2003 June 21 and 24 we obtained additional $BVRI$ images (Johnson-Cousins filter set), again with the 6.5-m Magellan/Clay telescope, this time using the MagIC camera (field of view of $2.4\arcmin \times 2.4\arcmin$ with a scale of $0.069\arcsec$/pixel; the seeing on both nights was approximately $0.6\arcsec$---see Figure~\ref{oc}).  Again, the frames were de-biased and flat-fielded with the IRAF package.  We derived astrometric solutions using 75 sources from the 2MASS catalog \citep{cutri01} and 49 sources from the USNO-B1.0 catalog.  The fit using the 2MASS sources is significantly better than the one using the USNO-B1.0 sources, and yields rms residuals of around 0\farcs07 in each coordinate.  From this, we derive an optical position of R.A.$ = \rm18^{h}13^{m}39\fs$04, Dec. = $-$33$^{\circ}46\arcmin$22\farcs3 (equinox J2000.0) with a 90\% confidence radius of 0\farcs2 \citep[the uncertainty is based on the astrometric accuracy of 2MASS, see e.g.][]{cutri01}.  This position is 0\farcs25 from the {\it Chandra}-derived X-ray position, well within the 0\farcs6 {\it Chandra} error circle.  This coincidence, the optical source's long-term as well as night-to-night variability, its blue color, and  the emission lines seen by \citet{steeghs03} argue strongly that it is the counterpart of \xtej. 

\begin{figure}
  \resizebox{!}{4in}{\includegraphics{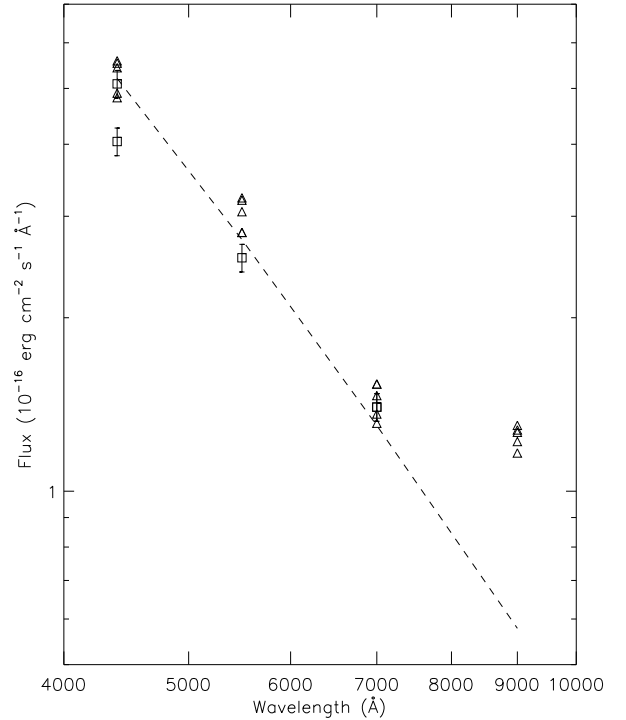}}
  	\caption{Optical data for \xtej.  Measurements taken on 2003 June 6 and 7 are plotted as squares, whereas those from the nights of 2003 June 21 and 24 are shown as triangles (errors are on the order of the symbol size).  The observed fluxes were dereddened assuming a Galactic extinction of $A_{V} = 0.71$. The emission predicted by an X-ray heated accretion disk model is plotted as a dashed line.  For this model, we used a source distance of 8 kpc with Galactic extinction, and minimized $\chi^{2}$ by setting cos $i = 0.6$.  Note that since these parameters are degenerate, the inclination angle fitted here is not well constrained.  The $I$-band flux (9000 \AA) lies well above the model prediction.}
	\label{opt_flux}
\end{figure}

The MagIC images were flux calibrated using the standard star Mark A2 \citep{landolt92}, and the magnitudes of the counterpart are presented in Table~\ref{mag}.  The counterpart on these later dates is on average a bit brighter than in the earlier observations, again agreeing with the observed X-ray flux of \xtej, which brightened from around 10 mCrab to around 12 mCrab before the second set of optical measurements were performed.  See Figure~\ref{opt_flux} for a plot of all optical data.

Finally, on 2004 March 15, while the source was in quiescence, we obtained an $R$-band image of the \xtej~field with the ESO Multi-Mode Instrument on the 3.5-m New Technology Telescope at the La Silla Observatory in Chile (field of view of $6.2\arcmin \times 6.2\arcmin$ with a scale of $0.167\arcsec$/pixel; the seeing was approximately $1.0\arcsec$).  The counterpart was not detected in this observation with a $3\sigma$ limiting magnitude of $R = 23.3$.

\section{X-ray Spectroscopy} \label{xspec}

We extracted separate X-ray spectra for the MEG and HEG data, co-added the $\pm$1 orders, and constructed the corresponding response files (ARFs and RMFs).  We used the {\tt contamarf}\footnote{\scriptsize{Information about contamarf can be found at http://space.mit.edu/CXC/analysis/ACIS\_Contam/ACIS\_Contam.html}} tool to correct the ARF for a decrease in low-energy sensitivity due to contamination on the ACIS CCDs \citep[see, e.g.,][]{marshall03}.  The count rate in the first-order MEG (HEG) spectrum was 2.9 (1.5) cts s$^{-1}$.  To improve statistics, we grouped the spectra such that there was a minimum of 100 counts per bin and assumed Poisson errors.  Background spectra were created using the script {\tt tg\_bkg} and subtracted prior to fitting.  All fitting was performed using the {\tt XSPEC} version 11.2 spectral analysis package.

Since the zeroth-order image is severely piled up, all spectral analysis was done on the dispersed first-order spectra, which are not affected by pileup.  We fit the MEG and HEG spectra simultaneously over the total energy range 0.5--10 keV, including a normalization factor which was allowed to vary between the two instruments.  We found the spectra to be best fit by an absorbed power-law plus blackbody model, where the equivalent hydrogen column density ($N_{\rm H}$) is fixed to the Galactic value of $1.63 \times 10^{21}$ cm$^{-2}$ \citep{dickey90}.  We note, however, that the flux in the blackbody component is only about 10\% of that in the power-law component.  We also fit an absorbed power-law with $N_{\rm H}$ allowed to vary, but the value did not deviate significantly from the Galactic value, justifying fits where this parameter is frozen.  The results of spectral fitting are summarized in Table~\ref{fit_param}.  Given that the fitted absorption is consistent with a column equal to the integrated Galactic value, it is likely that \xtej~lies at least 500 pc out of the Galactic plane.  At the Galactic latitude of $b$ = $-$7\fdg6, this implies a distance of $\gtrsim$ 3.8 kpc.  

\begin{deluxetable*}{ccccccccc}
\tabletypesize{\footnotesize}
\tablewidth{0pt}
\tablecaption{Spectral Fit Parameters\tablenotemark{a}}
\tablehead{
\colhead{$N_{\rm H}$}  & \multicolumn{3}{c}{Power-law}  & & \multicolumn{3}{c}{Blackbody}  &
 \colhead{} \\
  \cline{2-4}  \cline{6-8}  \\
\colhead{( $10^{21}$ cm$^{2}$)}   &  \colhead{$\Gamma$}  &  \colhead{$A_{1}$\tablenotemark{b}} & \colhead{Flux\tablenotemark{c}} & & \colhead{kT$_{\rm in}$ (keV)}  &   \colhead{$R_{\rm km}$\tablenotemark{d}}  & \colhead{Flux\tablenotemark{c}} &  \colhead{$\chi^{2}_{\nu}$ (dof)}}
\startdata
$1.67\pm0.17$ & $1.36\pm0.03$ & $3.71\pm0.15$ & 3.1 & & \nodata & \nodata & \nodata & 1.41 (405) \\
1.63 (fixed) & $1.35\pm0.02$ & $3.68\pm0.05$ & 3.1 & & \nodata & \nodata & \nodata & 1.41 (406) \\
1.63 (fixed) & $1.41\pm0.06$ & $3.32\pm0.11$ & 2.6 & & $0.95\pm0.13$ & $1.6\pm0.3$ & 0.3 & 1.29 (404) \\ 
\enddata 
\label{fit_param}
\tablenotetext{a}{All errors quoted are the 90\%-confidence range.}
\tablenotetext{b}{The amplitude of the power-law is the flux at 1 keV in units of $10^{-2}$ photons keV$^{-1}$ cm$^{-2}$ s$^{-1}$.}
\tablenotetext{c}{Fluxes are for the energy range 0.5--10 keV and are in units of 10$^{-10}$ erg cm$^{-2}$ s$^{-1}$.} 
\tablenotetext{d}{Blackbody radius assuming a distance to \xtej~of 8 kpc.}
\end{deluxetable*}

No significant spectral features (lines or edges) were observed.  To quantify this, we searched the spectral residuals for Gaussian features of FWHM equal to 800 km s$^{-1}$, to match the velocity seen in the H$\alpha$ emission \citep{steeghs03}.  The 3$\sigma$ upper limits are approximately 0.05 \AA~at 15 \AA~and 0.02 \AA~at 2.5 \AA.  To date, high-resolution X-ray spectroscopy has been obtained for four MSPs: XTE J1751$-$305 \citep{miller03}, XTE J0929$-$314 \citep{juett03}, XTE J1807$-$294 \citep{campana03}, and \xtej.  None have shown significant intrinsic spectral features, and the continua are generally well-fit by an absorbed power-law plus blackbody.  

\section{Discussion} \label{disc}

\begin{figure}
  \resizebox{!}{2.7in}{\includegraphics{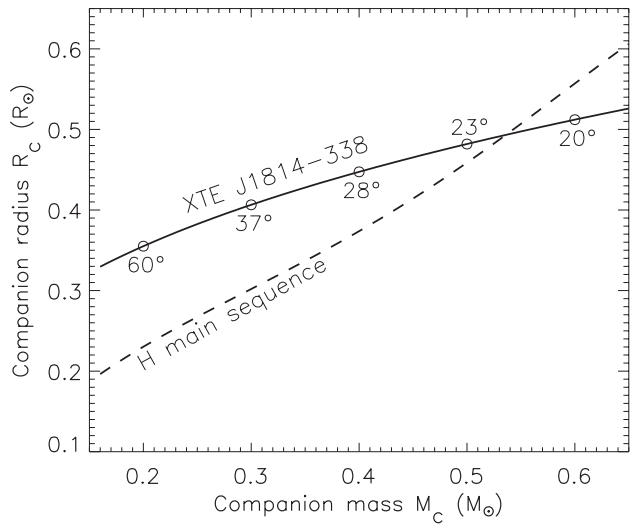}}
  	\caption{The mass-radius relationship for a Roche-lobe--filling companion of \xtej~(solid curve) and low-mass main-sequence stars (dashed curve) with corresponding inclination angles indicated.  The dashed curve is based on the analytic mass-radius function presented in \citet{tout96}.  The intersection of the curves suggests a companion mass of $\approx 0.5~M_{\sun}$.}
	\label{fig3}
\end{figure}

We have identified the optical counterpart of XTE J1814$-$334. The combined optical and X-ray observations allow us to place several constraints on the system parameters.  The mass-radius relation for a low-mass Roche-lobe--filling companion in a 4.27~hr binary is $R_c = 0.28\,(M_c/0.01 M_\sun)^{1/3}\,R_\sun$ \citep[see, e.g.,][]{frank02}.  Given the measured neutron star orbital parameters \citep{markwardt03c}, the lack of an X-ray eclipse thus restricts the binary inclination (defined as the angle between the line of sight and the orbital angular momentum vector) to $i< 77\degr$ ($\cos i>0.22$) for a Roche-lobe--filling companion.  In Figure~\ref{fig3}, we compare the companion's mass-radius relation with the theoretical relation for low-mass hydrogen main sequence stars \citep{tout96}. An ordinary hydrogen-rich companion is consistent with the required relation for a mass $M_c\approx 0.54 M_\sun$ and $i\approx 21\degr$ ($\cos i\approx 0.93$), although this value has a small a priori probability of 7\% for an isotropic sample of binary inclinations.  We note, however, that bloating of a hydrogen-rich companion owing to X-ray heating could allow a somewhat less massive companion to fill its Roche lobe \citep{tout89}, permitting a slightly larger inclination angle with a higher a priori probability.  A hydrogen main sequence companion is also consistent with the observed H and He lines in the optical spectra \citep{steeghs03}, whereas a white dwarf companion can be excluded as too small to fill its Roche lobe for any plausible donor mass.

We can check the consistency of our above methods using the brightness of the optical counterpart.  
\citet{shahbaz98} found that the magnitude of the optical outburst and the orbital period of soft X-ray transients are strongly correlated.  Using the relation they derive 
gives $\Delta V = 9.5 \pm 1.1$ for \xtej, and, since the counterpart in outburst is $V \approx 18.4$, a quiescent magnitude of $V_{\rm q} \approx 27.9$.  Assuming $A_V \geq 0.7$ due to Galactic dust along the line of sight \citep{schlegel98}, and a minimum distance of 6.4 kpc \citep[corresponding to the lower limit of the distance estimate derived from the radius expansion burst, see][]{strohmayer03}, we estimate a limit on the absolute magnitude of a main-sequence companion of $M_{V} < 13.2$.   We used the empirical mass-luminosity relation presented in \citet{delfosse00} to determine a minimum companion mass of $\approx 0.2~M_{\sun}$ (corresponding to $i < 60\degr$ and $\cos i \gtrsim 0.50$).  This companion mass is significantly lower than the value derived from the period-density relationship, and would imply that the companion star is significantly bloated due to X-ray heating.  However, we note that increasing the assumed distance to \xtej~will increase the minimum companion mass and decrease the maximum inclination angle.  

It is also instructive to consider what we can infer about the system given the $R > 23.3$ magnitude limit of the companion in quiescence.  We again assume $A_V = 0.7$, and infer a value of $A_R = 0.5$ \citep[given $A_V/E(B-V)=3.1$; see, e.g.,][]{cardelli89}.  We use the upper limit on the distance derived from the radius expansion burst of 9.6 kpc to give an absolute magnitude of $M_{R} = 7.9$.  This corresponds to a main-sequence spectral type of later than M1 V and a mass of $M \lesssim 0.5~M_{\sun}$ (corresponding to $i > 22\degr$ and $\cos i \lesssim 0.93$).  Using the harmonic properties of the burst oscillations, \citet{bhattacharyya05} were able to derive 90\% confidence intervals of $26\degr < i < 50\degr$ ($0.90 > \cos i > 0.64$).  This range is in agreement with our determination of the inclination angle from the magnitude limit of the optical counterpart ($i > 22\degr$) and lack of X-ray eclipse ($i < 77\degr$).  However, their lower limit of $i > 26\degr$ is a bit higher than the value we calculate given the mass-radius relation for an ordinary main-sequence companion of $21 \degr$, again suggesting that the companion may be bloated as a result of X-ray heating.  

We would expect the optical emission from \xtej~to originate from the combination of an X-ray--heated accretion disk and companion.  A simple X-ray--heated accretion disk model \citep{vrtilek90,chakrabarty98a,wang01} is able to account for the observed $BVR$ magnitudes for a wide range of plausible parameters, although the lack of data at bluer wavelengths precludes a well-constrained fit.  However, this model is unable to account for the $I$-band data, which are systematically brighter than predicted.
(For a plot of the optical data with a representative X-ray--heated accretion disk model, see Figure~\ref{opt_flux}.)  Although we do not have infrared data for \xtej, the sharp increase in flux in the $I$-band data suggests an IR excess with respect to the X-ray--heated disk model, similar to what was seen in the accretion-powered millisecond pulsar SAX J1808.4--3658 \citep{wang01}.  First, let us consider the possibility that this emission arises from the companion: if we take its $R$-band magnitude to be $R = 23.3$ (the limiting magnitude for its non-detection in quiescence), and assume it to be an M-dwarf (in agreement with previous mass estimates), this corresponds to an $I$-band magnitude of $I \sim 21.8$ \citep[again assuming $A_{V} = 0.71$ to obtain $A_{I} = 0.34$; see, e.g.,][]{cardelli89}.  Although the companion could brighten due to X-ray heating during the outburst, in order for it to account for the excess flux, it would have to brighten to $I \sim 18$, corresponding to a more than 30-fold increase in flux from quiescence.  We consider this to be highly unlikely.  Furthermore, if the secondary were substantially heated, its surface brightness would be highly anisotropic, and we would expect to see $I$-band variability at the timescale of the orbital period.  The 4.3-hour orbit is well-sampled on 21 June 2003, but we note that there is actually less variability at longer wavelengths (the percent RMS in $B$, $V$, $R$, and $I$ is 11\%, 9\%, 6\% and 4\%,  respectively.)  An alternative possibility is that the $I$-band excess could originate from synchrotron emission related to the outburst \citep[see, e.g.,][]{fender01}.  Future observations of \xtej~in outburst spanning a large spectral range, including measurements in the visible as well as at IR and radio wavelengths, would further our understanding of the geometry and emission mechanisms of \xtej.

\acknowledgments{We thank Harvey Tananbaum for providing us with Director's Discretionary time on the {\it Chandra} X-ray Observatory.  We thank the mission planning team for coordinating and executing a rapid response to our target of opportunity request, and Joy Nichols for processing and checking the data extremely quickly.  We thank Cole Miller for useful discussions, Jake Hartman for his assistance and helpful comments, and the anonymous referee for useful comments.  DS acknowledges support through a Smithsonian Astrophysical Observatory Clay Fellowship. PGJ is supported by NASA through Chandra Postdoctoral Fellowship grant number PF3--40027 awarded by the Chandra X--ray Center, which is operated by the Smithsonian Astrophysical Observatory for NASA under contract NAS8--39073. This work was supported in part by NASA under contract NAS 8-01129.}

\end{document}